\documentclass[twocolumn,amssymb,nobibnotes,showpacs,superscriptaddress,aps,reprint]{revtex4-1}
\usepackage{graphicx,amsmath,natbib,bm}

\allowdisplaybreaks

\begin{document}

\title{Electromagnetic control of the collective radiations with three-photon blockade}

\author{Y. F. Han}
%\email{hanyufeng@tongji.edu.cn}
\affiliation{MOE Key Laboratory of Advanced Micro-Structured Materials, School of Physics Science and Engineering, Tongji University, Shanghai 200092, China}
\affiliation{School of Mathematics and Physics, Anhui University of Technology, Ma'anshan 243032, China}

\author{C. J. Zhu}
\email[]{cjzhu@tongji.edu.cn}
%\thanks{}
\affiliation{MOE Key Laboratory of Advanced Micro-Structured Materials, School of Physics Science and Engineering, Tongji University, Shanghai 200092, China}

\author{J. P. Xu}
%\email[]{cjzhu@tongji.edu.cn}
\affiliation{MOE Key Laboratory of Advanced Micro-Structured Materials, School of Physics Science and Engineering, Tongji University, Shanghai 200092, China}

\author{Y. P. Yang}
\email[]{yang\_yaping@tongji.edu.cn}
%\thanks{}
\affiliation{MOE Key Laboratory of Advanced Micro-Structured Materials, School of Physics Science and Engineering, Tongji University, Shanghai 200092, China}


%\author{G. S. Agarwal}
%\email[]{girish.agarwal@tamu.edu}
%\affiliation{Institute for Quantum  Science and Engineering,
%	and Department of Biological and Agricultural Engineering
%	Texas A\&M University, College Station, TX 77843, USA}
%\affiliation{Department of Physics, Oklahoma State University, Stillwater, Oklahoma 74078, USA}

\date{\today}

\begin{abstract}
We study properties of collective radiations of coherently driven two three-level ladder-type atoms trapped in a single-mode cavity. Using the electromagnetically induced transparency technique, we show that the three-photon blockade effect can be observed and the properties of collective radiations are strongly dependent on the phase between two atoms. In the case of in-phase radiations, the frequency range to realize the three-photon blockade can be broadened as the control field increases. However, in the regime of the three-photon blockade, the property of collective radiations changes from hyperradiance to subradiance. In the case of out-of-phase radiations, hyperradiance accompanied with the three-photon blockade can be observed. The results presented in this paper show that our scheme is an attractive candidate to generate antibunched photon pairs and control the properties of collective radiations.
\end{abstract}


\maketitle

%\section{INTRODUCTION}
%
\section{Introduction}
As one of the most fascinating topics in the field of quantum optics, superradiance, and subradiance have arisen extensive attention in both theory and experiments since its discovery by Dicke~\cite{dicke1954coherence,rehler1971superradiance,scully2006directed,akkermans2008photon,rohlsberger2010collective,monz201114,svidzinsky2013quantum,feng2014effect,scully2015single,longo2016tailoring}. In earlier years, Agarwal et al. ascribed the physical mechanism of the superradiant collective emission to strong quantum correlations among atoms lying in symmetric Dicke states~\cite{agarwal1974quantum,gross1982superradiance}. Later, it is found that the superradiance or subradiance results from the multiparticle entanglement of the Dicke state via the interference of quantum pathways~\cite{wiegner2011quantum,nienhuis1987spontaneous}. The experimental works by Blatt et al. also show that the entanglement plays an indispensable role in the collective emission of radiation~\cite{monz201114,devoe1996observation}.

In recent years, Scully et al.~\cite{scully2015single} theoretically studied the dynamics of single photon superradiance and subradiance, and R\"{o}hlsberger et al.~\cite{rohlsberger2010collective} carried out an experimental work by embedding an ensemble of resonant atoms in the center of a planar cavity. Due to the back reaction in cavity QED systems, many novel features of the collective radiations have been studied theoretically and demonstrated experimentally~\cite{friedberg1973frequency,agarwal1974quantum,gross1979maser,rempe1990observation,reimann2015cavity}, which help us to get a further understanding of superradiant or subradiant collective emission. In particular, Pleinert et al. predict the possibility of hyperradiance arising from the collective radiations in a strongly coupled two atoms cavity QED system~\cite{pleinert2017hyperradiance}. As the hyperradiance occurs, the mean photon number can be up to $2$ orders of magnitude larger than the case of a single atom. Xu et al.~\cite{Xu2017Hyperradiance} show that the hyperradiance can also be observed when the pump field drives two atoms under the non-resonant condition~\cite{agarwal1984vacuum,raimond2001manipulating}.

Compared with the cavity driving schemes, which is generally considered in cavity QED systems, the atom driving enhances the optical nonlinearity so that many interesting quantum and nonlinear features of cavity photons can be observed, including the improvement of two-photon blockade~\cite{zhu2017collective}, observation of three-photon blockade~\cite{hamsen2017two}, and realization of hyper-radiance phenomenon~\cite{pleinert2017hyperradiance,Xu2017Hyperradiance,pleinert2018phase}. In particular, we show that the asymmetry in the atom-cavity coupling strengths open new pathways for multiphoton blockade which result in the three-photon blockade phenomenon with reasonable mean photon number over a broad frequency regime.

Combining the electromagnetically induced transparency technique with the two atoms cavity QED system, In this paper, we show that the collective radiation properties are strongly dependent on the control field Rabi frequency and the phase shift between two atoms. It is shown that the three-photon blockade can be observed not only under the condition of out-of-phase radiations but also in-phase radiations. The frequency range for achieving the three-photon blockade can be broadened by increasing the control field Rabi frequency. In particular, the property of collective radiations changes from subradiance to hyperradiance accompanying with the three-photon blockade effect. Therefore, it is possible to generate photon pairs efficiently in this system.

\section{Model}
%
% fig: 1 %%%%%%%%%%%%%%%%%%%%%%%%
\begin{figure}[htb]
	\centering
	\includegraphics[width=0.5\textwidth]{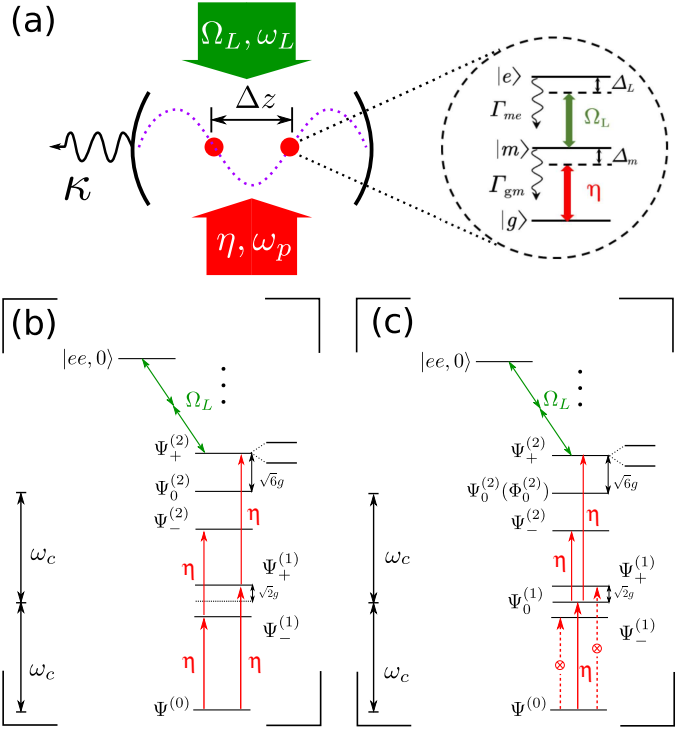}
	\caption{(a) The sketch of the two atoms cavity QED system, where two three-level ladder-type atoms are trapped in a single mode cavity. The energy levels are labeled as $|\alpha\rangle\ (\alpha=g,m,e)$ with energy $\hbar\omega_\alpha$. A pump (control) field $\eta$ ($\Omega_L$) with angular frequency $\omega_p$ ($\omega_L$) couples the $|g\rangle\leftrightarrow|m\rangle$ ($|m\rangle\leftrightarrow|e\rangle$) transition. Here, $\Gamma_{me}$ ($\Gamma_{gm}$) is the spontaneous decay rate of state $|e\rangle$ ($|m\rangle$), and the cavity decay rate is $\kappa$. $\Delta_\alpha$ is the detuning of state $|\alpha\rangle$. (b) and (c) show the dressed state pictures of $\phi_z=0$ and $\pi$, respectively.}
	\label{fig:fig1}
\end{figure}
%%fig: 1%%%%%%%%%%%%%%%%%%%%%%%%%
The configuration of this two atoms cavity QED system is shown in Fig.~1(a), where two identical three-level ladder-type atoms are trapped in a single mode cavity at different positions $z_i$ ($i=1,2$). For each atom, the energy levels are labeled as $|g\rangle$, $|m\rangle$ and $|e\rangle$, respectively. As shown in Fig.~1(a), a weak pump field with angular frequency $\omega_p$ drives the $|g\rangle\leftrightarrow|m\rangle$ transition, but a strong control field with angular frequency $\omega_L$ couples $|m\rangle\leftrightarrow|e\rangle$ transition. In our system, we assume that the cavity mode with resonant frequency $\omega_{\rm cav}=2\pi c/\lambda_{\rm cav}$ only couples the $|g\rangle\leftrightarrow|m\rangle$ transition. The position-dependent coupling strengths between the atom and cavity are then given by $g_i=g\cos{(2\pi z_i/\lambda_{\rm cav})}$ where $z_i$ is the position of the $i$-th atom ($i=1,2$), $g$ is the maximum atom-cavity coupling strength and $\lambda_{\rm cav}$ is the wavelength of the cavity mode. For mathematic simplicity, we also assume that one atom is fixed at the antinode of the cavity (i.e., $z_1=0$), while the position of the other atom can be varied along the axis of the cavity. Therefore, the atom-cavity coupling strengths are given by $g_1=g$ and $g_2=g\cos{(\phi_z)}$ with $\phi_z=2\pi\Delta z/\lambda_{\rm cav}=2\pi(z_2-z_1)/\lambda_{\rm cav}$.

Under the electric dipole and rotating wave approximations, the dynamical behavior of the whole system shown in Fig.~1(a) can be treated by a master equation given by
\begin{equation}
\frac{d}{dt}\rho=-\frac{i}{\hbar}[H,\rho]+{\cal L}_{\kappa}\rho+{\cal L}_{\Gamma}\rho,
\label{eq:master}
\end{equation}
where $\rho$ is the density operator of the atom-cavity system. The Hamiltonian of the whole system $H=H_{0}+H_{I}+H_{D}$, where $H_0$, $H_I$ and $H_D$ denotes the energy of the atoms and cavity, the interaction between the atom and cavity, and the driving terms, respectively. In our system, we have  $H_{0}=\hbar\sum_{j=1}^2(\Delta_{m}S^{j}_{mg}S^{j}_{gm}+\Delta_{e}S^{j}_{em}S^{j}_{me})+\hbar\Delta_{\rm cav}a^{\dag} a$, $H_{I}=\hbar\sum_{j=1}^2[g_{j}(aS^{j}_{mg}+a^{\dag} S^{j}_{gm})+\Omega_{L}(S^{j}_{em}+S^{j}_{me})]$ and $H_{D}=\hbar \eta\sum_{j=1}^2(S^{j}_{mg}+S^{j}_{gm})$. Here, $\Omega_L$ and $\eta$ are the Rabi frequency of the control field and pump field, respectively.  $S^j_{\alpha\beta}=|\alpha\rangle_j\langle\beta|\ (\alpha,\beta={g,m,e})$ is the atomic raising and lowering operator of the $j$-th atom ($j=1,2$). $a$ and $a^{\dag}$ are the annihilation and creation operator of the cavity mode. The detunings are defined as $\Delta_{m}=\omega_{m}-\omega_g-\omega_{p}$, $\Delta_{e}=\omega_{e}-\omega_g-(\omega_p+\omega_c)=\Delta_m+\Delta_{L}$ with $\Delta_L=\omega_e-\omega_m-\omega_L$ and $\Delta_{\rm cav}=\omega_{\rm cav}-\omega_{p}$, respectively. It is worth to point out that, to avoid the dipole-dipole interactions~\cite{goldstein1997dipole}, the separation of two atoms must be larger than the cavity wavelength $\lambda_{\rm cav}$.

The spontaneous decays of the atomic states are introduced by the Liouvillian operators, i.e.,  ${\cal L}_{\Gamma}\rho=\sum_{j=1}^2[\Gamma_{gm}(2S_{gm}^{j}\rho S_{mg}^{j}-S_{mg}^{j}S_{gm}^{j}\rho-\rho S_{mg}^{j}S_{gm}^{j})+\Gamma_{me}(2S_{me}^{j}\rho S_{em}^{j}-S_{em}^{j}S_{me}^{j}\rho-\rho S_{em}^{j}S_{me}^{j})]$, where $\Gamma_{\alpha\beta}$ denotes the spontaneous decay rate from state $|\beta\rangle$ to state $|\alpha\rangle$. The Liouvillian term describing the cavity decay at rate $\kappa$ is given by ${\cal L}_{\kappa}\rho=\kappa(2a\rho a^{\dag}-a^{\dag} a \rho-\rho a^{\dag} a)$. Solving Eq.~(1) numerically, one can examine the features of the collective radiations and the quantum properties of the cavity field simultaneously.

In general, the quantum properties of an optical field can be characterized by the field correlation function, which can also be used to describe the statistical properties of the fields, such as bunching and anti-bunching behaviors. In quantum field theory, the steady-state second-order and third-order field correlation functions are defined as
\begin{equation}
  g_{\rm ss}^{(2)}(0)=\frac{\langle a^{\dag} a^{\dag} a a \rangle}{\langle a^{\dag} a \rangle^{2}},
\end{equation}
and
\begin{equation}
  g_{\rm ss}^{(3)}(0)=\frac{\langle a^{\dag} a^{\dag} a^{\dag} a a a \rangle}{\langle a^{\dag} a \rangle^{3}},
\end{equation}
respectively. Here, $g^{(2)}_{\rm ss}(0)>1$ denotes the bunched photons, whereas $g^{(2)}_{\rm ss}(0)<1$ denotes the anti-bunched photons. Indeed, $g^{(2)}_{\rm ss}(0)<1$ is an important witness for the two-photon blockade effect and is used to characterize  the quality of single-photon sources in applications. $g^{(3)}_{\rm ss}(0)$ is the third-order field correlation function, which is used to characterize the probability of simultaneous arrival of three photons. If $g^{(3)}_{\rm ss}(0)<1$ but $g^{(2)}_{\rm ss}(0)>1$, one can observe that two photons arrive together, but the third photon arrives at a different time. This condition implies the three-photon blockade effect and can be used to characterize the quality of two-photon sources~\cite{zhu2017collective}.

\begin{figure*}[htbp]
\centering
	\includegraphics[width=\textwidth]{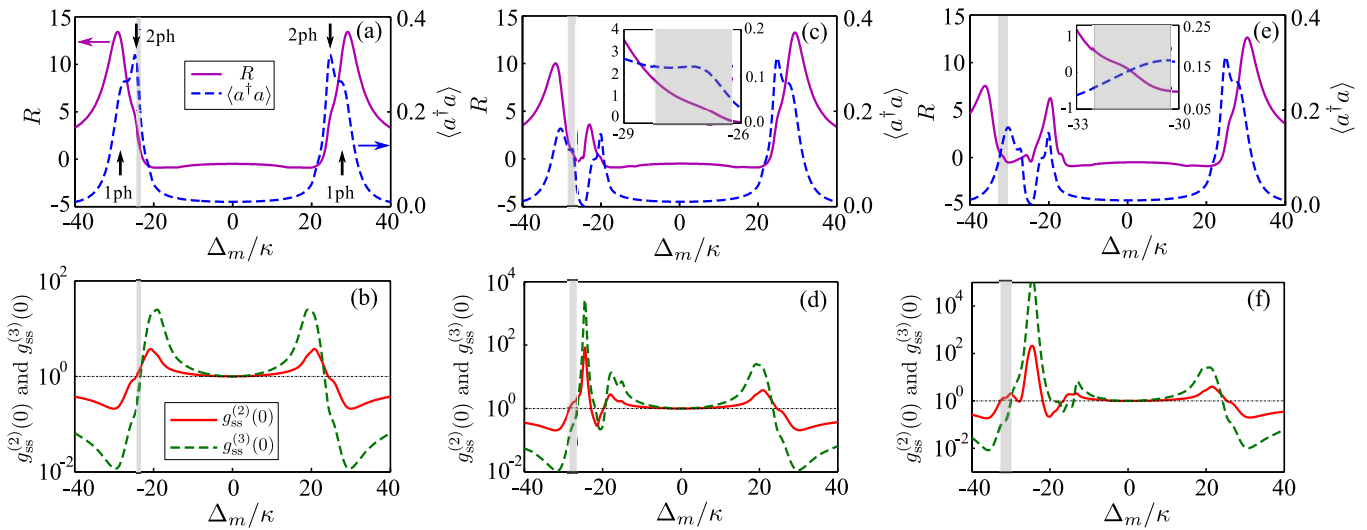}
	\caption{Panels (a,c,e) show the parameter $R$ (violet solid curve) and mean photon number $\langle a^{\dag}a \rangle$ (blue dashed curve). Panels (b,d,f) show the field correlation functions $g^{(2)}_{ss}(0)$ (red solid curve) and $g^{(3)}_{ss}(0)$ (green dashed curve). The gray areas indicate the frequency range to realize the three-photon blockade, and the horizontal dotted lines indicate $g^{(2)}_{ss}(0)=g^{(3)}_{ss}(0)=1$. Here, we choose $\phi_z=0$ and the control field Rabi frequency $\Omega_L=0$ [panels (a,b)], $4.8\kappa$ [panels (c,d)] and $11.0\kappa$ [panels (e,f)], respectively. Other system parameters are given in the content.}
\label{fig:fig2}
\end{figure*}
On the other hand, the behavior of collective radiations can be characterized by a witness parameter $R$, which is given by~\cite{pleinert2017hyperradiance},
\begin{equation}\label{eq:R}
 R=\frac{\langle a^{\dag} a \rangle _{2}-2\langle a^{\dag} a \rangle _{1}}{2\langle a^{\dag} a \rangle _{1}}
\end{equation}
Here, $\langle a^{\dag} a \rangle _{2}$ is the mean photon number in two atoms cavity QED system, and $\langle a^{\dag} a \rangle _{1}$ is the mean photon number in a single atom-cavity QED system. Obviously, $R<0$ means that the collective radiations are suppressed (i.e., subradiance), and $R=0$ indicates that the radiation rate is the same as that of the single atom case. However, $R>0$ shows that the collective radiations are enhanced compared with the single atom case. Particularly, $R=1$ (i.e., $\langle a^{\dag} a \rangle _{2}=4\langle a^{\dag} a \rangle _{1}$) indicates that the radiation rate scales with the square of the number of atoms $\propto N^{2}$, corresponding to the superradiance phenomenon~\cite{auffeves2011few}. Furthermore, $R>1$ denotes that the collective radiations are significantly enhanced, which is defined as the  hyperradiance~\cite{pleinert2017hyperradiance}.

\section{In-phase radiations}
We first consider the case that two atoms radiate in phase (i.e., $\phi_z=0$), where $g_{1}=g_{2}=g$. In the absence of the control field, both theoretical and experimental works show that the three-photon blockade is hard to be observed~\cite{hamsen2017two,zhu2017collective}. As shown in Fig.~2(a), there exist four peaks in the cavity excitation spectrum, corresponding to the one-photon and two-photon excitations, respectively. In a very narrow frequency regime near the two-photon excitation, the three-photon blockade can be achieved with the superradiance and even hyper-radiance behavior [$R\geq1$, see the gray area in Fig.~2(a) and (b)]. Here, the system parameters are given by $g=20\kappa,\Gamma_{gm}=\kappa,\Gamma_{me}=\kappa/100,\eta=2\kappa$\cite{pritchard2010cooperative,petrosyan2011electromagnetically}. In the presence of the control field, the left side peak in the cavity excitation spectrum is split into two peaks due to the coupling of the control field [see Fig.~2(c)]. As a result, the frequency range to realize the three-photon blockade is broadened and the collective radiation is still enhanced ($0<R<1$) as shown in Fig.~2(c) and (d) (gray areas). Further increasing the control field Rabi frequency, although the behavior of the collective radiations changes from superradiance to subradiance ($R<0$), the three-photon blockade can still be observed over a wide frequency range [see Fig.~2(e) and 2(f)].

The physical mechanism can be explained by exploring the eigenstates of this multilevel system. Assuming that the pump and control fields are not very strong and can be treated as perturbations of the system, we first consider the interaction between the cavity field and states $|g\rangle$ and $|m\rangle$. In this case, the eigenvalues and eigenstates can be obtained easily by using the collective states $|gg\rangle$, $|\pm\rangle=(|mg\rangle\pm|gm\rangle)/\sqrt{2}$ and $|mm\rangle$ to rewrite the Hamiltonian~\cite{pleinert2017hyperradiance,zhu2017collective,han2018electromagnetic}. As shown in Fig.~1(b), we have the eigenstates $\Psi_{\pm}^{(1)}=(\pm|gg,1\rangle+|+,0\rangle)/\sqrt{2}$ with eigenvalues $E_{\pm}^{(1)}=\hbar\omega_{\rm cav}\pm\sqrt{2}g\hbar$ in one-photon space. In two-photon space, the eigenstates are given by  $\Psi_{\pm}^{(2)}=|gg,2\rangle/\sqrt{3}\pm|+,1\rangle/\sqrt{2}+|mm,0\rangle/\sqrt{6}$ with eigenvalues $E_{\pm}^{(2)}=2\hbar\omega_{\rm cav}\pm\sqrt{6}g\hbar$ and $\Psi_{0}^{(2)}=(-\sqrt{3}|gg,2\rangle+\sqrt{6}|mm,0\rangle)/3$ with eigenvalues $E_{0}^{(2)}=2\hbar\omega_{\rm cav}$. Taking $\Delta_{L}=\sqrt{6}g/2$, we find that the control field couples the $\Psi_{+}^{(2)}\rightarrow|ee,0\rangle$ transition resonantly via two-photon process. Therefore, the state $\Psi_{+}^{(2)}$ is split into a doublet, corresponding to two left side peaks shown in Fig.~2(c) and 2(e).

\begin{figure*}[htbp]
\centering
	\includegraphics[width=\textwidth]{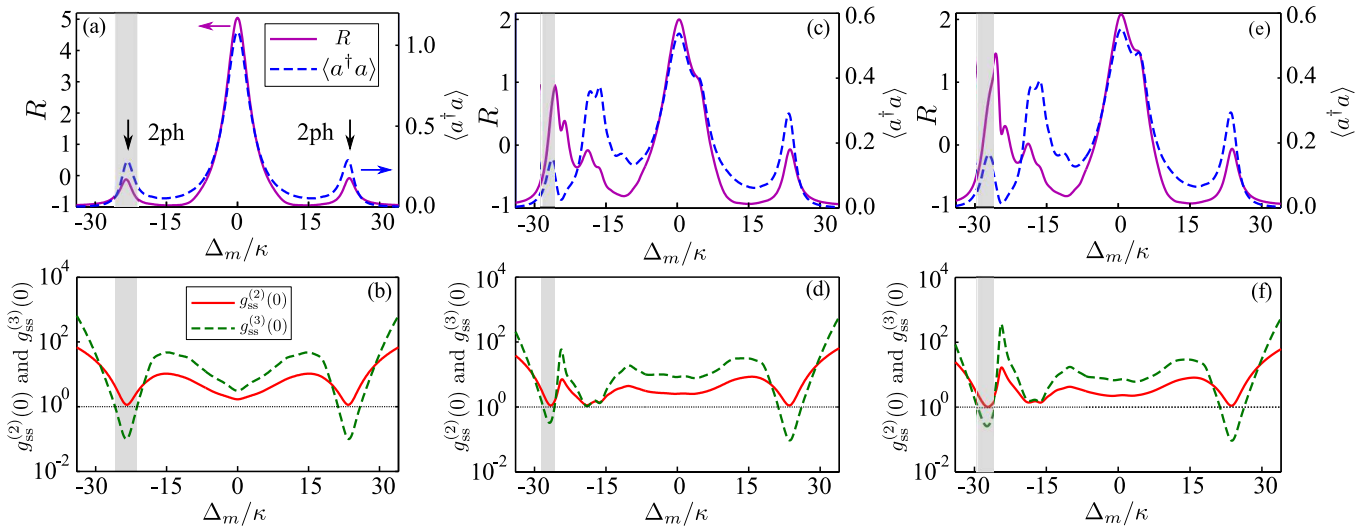}
	\caption{Panels (a,c,e) show the parameter $R$ (violet solid curve) and mean photon number $\langle a^{\dag}a \rangle$ (blue dashed curve). Panels (b,d,f) show the field correlation functions $g^{(2)}_{ss}(0)$ (red solid curve) and $g^{(3)}_{ss}(0)$ (green dashed curve). The gray areas indicate the frequency range to realize the three-photon blockade, and the horizontal dotted lines indicate $g^{(2)}_{ss}(0)=g^{(3)}_{ss}(0)=1$. Here, we choose $\phi_z=\pi$ and the control field Rabi frequency $\Omega_L=0$ [panels (a,b)], $4.0\kappa$ [panels (c,d)] and $5.6\kappa$ [panels (e,f)], respectively. Other system parameters are the same as those used in Fig.~2, except for $\eta=6\kappa$.}
\label{fig:fig3}
\end{figure*}
\section{Out-of-phase radiations}
Now, we consider that two atoms radiate out-of-phase, i.e., $\phi_{z}=\pi$ yielding  $g_{1}=-g_{2}=g$. Neglecting the pump and control field, one can obtain three eigenstates in one-photon space, labeled as $\Psi_{\pm}^{(1)}=(\pm|gg,1\rangle+|-,0\rangle)/\sqrt{2}$ with eigenvalues $E_{\pm}^{(1)}=\hbar\omega_{\rm cav}\pm \sqrt{2}g\hbar$ and $\Psi_{0}^{(1)}=|+,0\rangle$ with eigenvalues $E_{0}^{(1)}=\hbar\omega_{\rm cav}$, respectively. In two-photon space, there exist four eigenstates, i.e., $\Psi_{\pm}^{(2)}=-|gg,2\rangle/\sqrt{3}\mp|-,1\rangle/\sqrt{2}+|mm,0\rangle/\sqrt{6}$ with eigenvalues $E_{\pm}^{(2)}=2\hbar\omega_{\rm cav}\pm \sqrt{6}g\hbar$, two degenerate states  $\Psi_{0}^{(2)}=|gg,2\rangle/\sqrt{3}+\sqrt{6}|mm,0\rangle/3$ and $\Phi_{0}^{(2)}=|+,1\rangle$ with the same eigenvalue $E_{0}^{(2)}=2\hbar\omega_{\rm cav}$, as depicted in Fig.~1(c). Different from the case of $\phi_z=0$, the asymmetry coupling strengths cause that the one-photon excitation pathways are forbidden~\cite{zhu2017collective}. Therefore, two-photon (two side peaks) and multiphoton (central peak) excitations are dominant and three-photon blockade effect can be observed over a wide frequency range as shown in Fig.~3(a) and (b). When the control field is tuned to be resonant with the $|ee,0\rangle\rightarrow \Psi^{(2)}_{+}$ transition via the two-photon process as shown in Fig.~1(c), the two-photon excitation state $\Psi^{(2)}_{+}$ will be split into a doublet so that one can observe four peaks in the cavity excitation spectrum [see Fig.~3(b)]. As shown in panels (d) and (f), the three-photon blockade can always be observed over a wide frequency range when the control field is turned on. However, the behavior of the collective radiation changes from superradiance to hyperraidance as the control field Rabi frequency increases [see panels (c) and (e)]. It must be pointed out that these features of the collective radiations and the quantum properties of the cavity field exhibited in out-of-phase radiation case can not be observed in single atom-cavity QED systems.

\section{Conclusion}
We have studied the behavior of collective radiations of cavity field under the condition of the three-photon blockade in a two atoms cavity QED system. By using the EIT technique, we show that the behavior of the collective radiation and quantum properties of the cavity field can be greatly changed by tuning the control field Rabi frequency. Even in the case of in phase radiations, the three photon blockade effect can be observed over a wide frequency range. Under the condition of three-photon blockade, more interestingly, the behavior of the collective radiations can be changed from superradiance to hyperraidance by just increasing the control field Rabi frequency in the case of out phase radiations. The results presented here shows that our scheme is a good candidate to generate photon pairs via the three-photon blockade effect.

\begin{acknowledgments}
We acknowledge the National Key Basic Research Special Foundation (Grant No. 2016YFA0302800); the Shanghai Science and Technology Committee (Grant No. 18JC1410900); the National Nature Science Foundation (Grants No. 11774262, 11474003, 11504003, 61675006).
\end{acknowledgments}

\bibliographystyle{apsrev4-1}
%\bibliography{supRadiance_hyf}% Produces the bibliography via BibTeX.
%\bibliographystyle{aipauth4-1}
%\bibliographystyle{apsrmp4-1}
\bibliography{supRadiance_hyf}% Produces the bibliography via BibTeX.

%\begin{thebibliography}{99}
%	%
%	
%\end{thebibliography}
%
\end{document}